\documentclass[prl,aps,floatfix,twocolumn]{revtex4-1}

\usepackage{amsmath,amssymb}
\usepackage{graphicx}
\usepackage{color}

\usepackage{verbatim}

\begin{document}
%\draft

%\begin{comment}
\title{Thermo-diffusion in inertially confined plasmas}

\author{Grigory~Kagan\footnote{Email: kagan@lanl.gov} and Xian-Zhu Tang}
\affiliation{Theoretical Division,
Los Alamos National Laboratory,
Los Alamos, NM 87545}
\date{\today}

\begin{abstract}
  In a plasma of multiple ion species, thermodynamic forces such as
  pressure and temperature gradients can drive ion species separation via
  inter-species diffusion. Unlike its neutral mix counterpart, plasma
  thermo-diffusion is found comparable to, or even much larger than,
  baro-diffusion. It is shown that such a strong effect is due to the
  long-range nature of the Coulomb potential, as opposed to
  short-range interactions  in neutral gases.  A special composition of the 
   tritium and \textsuperscript{3}He fuel is
  identified to have vanishing net diffusion during adiabatic compression,
  and hence provides an experimental test in which yield degradation
  is minimized during ICF implosions.
 \end{abstract}

\maketitle
%\end{comment}

%\tableofcontents

In inertial confinement fusion (ICF), the central ``hot-spot" plasma,
assembled by laser-driven spherical implosion~\cite{ICF}, contains
multiple ion species.  Common combinations include low-Z fuel mixtures such as deuterium
(D)/tritium (T) and D\textsuperscript{3}He with possible addition of high-Z pusher
ions, such as carbon or silicon, due to plastic or glass
 shell mixing~\cite{C-K-Li-mix, wilson-mix-2011} into the gas fill. Sometimes high-Z gas
dopants such as Ar~\cite{regan-Ar} or Kr~\cite{yaakobi-Kr} are
intentionally introduced for diagnostic purposes, as well as
to specifically study the pre-mix effects~\cite{wilson-premix,
  garbett-premix}. The powerful thermodynamic forces (e.g. pressure~\cite{amendt-baro, zeldovich, landau-hydro}
and temperature gradients~\cite{zeldovich, landau-hydro}) in an imploding target can drive ion
species separation via inter-species diffusion.  Observation of the resulting fuel stratification in the DT implosion,
which upsets the initially optimal arrangement of equal number
densities of D and T, in experiments~\cite{casey-prl} and 
kinetic simulations~\cite{Larroche-2012, bellei-2013, Larroche-comment, bellei-reply-to-Larroche} have recently been reported.
The targets with high-Z dopants show a particularly strong yield
anomaly~\cite{wilson-premix, garbett-premix, dodd-laser-absorp},
suggesting that even stronger fuel stratification may take place. 

Perhaps the most intriguing physics aspect of inter-ion-species diffusion in
a collisional plasma is the role of thermo-diffusion, which, as its
name suggests, is driven by the gradients of ion and electron
temperatures.  The novelty comes through as a sharp contrast to the
better-known case of a neutral mixture, where thermo-diffusion is
substantially less important than baro-diffusion, though often 
counteracts it~\cite{zeldovich}. According to statistical
physics,  thermo-diffusion strongly depends
on the details of the collisional exchange between and within the
species~\cite{landau-hydro}. Due to the long range nature of Coulomb
collisions in plasmas, as opposed to short range collisions between
neutral particles, one may expect thermo-diffusion in plasmas and
neutral mixtures to be fundamentally different. This difference becomes particularly striking with the observation that
 plasma baro-diffusion ratio $k_p$ is identical
to its neutral counterpart~\cite{electro-diffusion}.

The purpose of this Letter is to present (i) an intuitively obvious,
qualitative analysis that elucidates the enhanced importance of thermo-diffusion
in a plasma in comparison with a neutral mixture; (ii) a
first-principle derivation of the total diffusive mass flux in a
plasma with two ion species.  Specifically for (ii), thermo-diffusion
ratios associated with the ion and electron temperature gradients, as
well as classical diffusion coefficients, are calculated numerically
for selected pairs of ion species. In contrast to the neutral mixture
case, contributions of thermo- and baro-diffusions are found
comparable and reinforcing each other. Even more importantly, when the
charge number of one of the ion species becomes large, the former
dominates over the latter. The newly developed theory for inter-ion-species diffusion is
then employed to investigate whether fuel stratification can be
mitigated during ICF implosion.  Unlike the prediction of
Ref.~\cite{amendt-lapse-rate}, which implicitly relies on thermo-diffusion being zero~\cite{amendt-lapse-rate-reply}, 
suppressing the instantaneous diffusive flux 
is found   possible only in very special mixtures such as
T\textsuperscript{3}He, where electro-diffusion is strong enough to
counteract thermo-diffusion.

{\bf {Inter-ion-species diffusion}}: 
Following Landau and
Lifshitz~\cite{landau-hydro}, the diffusive mass flux in a plasma with
two ion species is defined as
\begin{equation}
\label{eq: flux-def}
{\bf i} = \rho_{l}( {\bf u}_{l}-{\bf u}),
\end{equation}
where $\ {\bf u} \equiv (\rho_{l} {\bf u}_{l} + \rho_{h} {\bf
  u}_{h})/\rho$ is the center-of-mass velocity with $\rho_{\alpha}$, $
{\bf u}_{\alpha}$ the mass density and fluid velocity of species
``$\alpha$", respectively, and subscripts ``$l$" and ``$h$" refer to
the light and heavy ion species. Electron inertia is negligible, so
$\rho \equiv \rho_l + \rho_h$ can be referred to as the plasma density.
The diffusive flux governs the evolution of
the mass concentration $c \equiv \rho_l/\rho$ through~\cite{landau-hydro}
\begin{equation}
\label{eq:dcdt}
\rho \frac{\partial c}{\partial t} + \rho {\bf u}\cdot\nabla c + \nabla\cdot{\bf i} = 0.
\end{equation}

Allowing different electron and ion temperatures ($T_e$ and $T_i$), it was recently
shown~\cite{electro-diffusion} that the diffusive mass flux in a plasma with two ion species
take the form
\begin{widetext}
\begin{equation}
\label{eq: canonical-flux}
{\bf i} =
- \rho D \Bigl( \nabla c +k_p \nabla \log{p_i} + \frac{e k_E}{T_i}\nabla \Phi + k_T^{(i)} \nabla \log{T_i}  + k_T^{(e)} \nabla \log{T_e}\Bigr),
\end{equation}
\end{widetext}
where $p_i = p_l + p_h$ is the ion pressure. The terms on the right
side of Eq.~(\ref{eq: canonical-flux}) represent contributions from,
respectively, classical, baro-, electro- and
thermo-diffusions. Classical diffusion relaxes concentration
perturbations in the absence of externally imposed gradients. Physical
picture underlying the other diffusion mechanisms can be understood by
considering the relative motion of the light and heavy ion fluids due
to external forces. For example, baro-diffusion is due to the $\nabla
p_i$ force giving a larger acceleration to fluid elements of the light
ion species, making ${\bf u}_{l}$ and ${\bf u}_{h}$ differ. Similarly,
electro-diffusion is due to the electric field giving a larger
acceleration to the species with higher charge-to-mass ratio. Finally,
accelerations of fluid elements of the light and heavy ion species due
to the ion-ion and ion-electron thermal forces are generally different
as well, giving rise to thermo-diffusions proportional to $\nabla T_i$ and
$\nabla T_e$, respectively. The relative importance of
various diffusion mechanisms can therefore be understood by comparing
the corresponding forces.

{\bf Thermo-diffusion strongly enhanced due to the long range nature
  of Coulomb interaction -- a qualitative demonstration:} Before
presenting a rigorous first principle calculation of the various
diffusion coefficients in Eq.~(\ref{eq: canonical-flux}), we utilize
the intuitive picture of the preceding paragraph to elucidate the
importance of thermo-diffusion in plasmas, as opposed to the minor
role it plays in neutral gas mixtures.  We begin by adapting the
qualitative theory of Braginskii~\cite{braginskii} to estimate the
thermal force in a binary mixture with the collision frequency role explicitly retained.

Consider the two mixture components at a point $x_0$ with their net
flow velocities equal zero. There are two groups of light particles
(``$-$" and ``$+$") coming from $x_0 - \lambda$ and $x_0 + \lambda$,
respectively, where $\lambda$ is the mean free path. The thermal force
experienced by the light component as a whole arises due to the
frictional forces experienced by the two groups, $R_{-}$ and $R_{+},$
giving a non-vanishing superposition~\cite{braginskii}. These forces
are estimated as $R_{-,+} \sim \nu_{lh} m_l n_{-,+} v_{-,+}$, where
$m_l$ is the light particle mass and $\nu_{lh}$ is the characteristic
frequency of collisions between the species. $n_{-,+}$ and $v_{-,+}$
denote the densities and the net flow velocities of the corresponding
groups of light particles; for the overall light fluid to remain at
rest they must satisfy $n_{-} v_{-} + n_{+} v_{+} = 0$. Expanding $T$
about $x_0$ and noticing that $n_{-,+} v_{-,+} \sim n_l v_{th}$ with
$n_l$ and $v_{th}$ being the number density and thermal velocity of
light particles, respectively, the resultant force is estimated as
\begin{align}
R_T & = R_+ - R_- \sim
m_l n_l v_{th} \times 2\lambda \frac{\partial \nu_{lh}}{\partial T} \frac{d T}{d x} \nonumber \\
& \propto
\frac{\partial \ln{ \nu_{lh}}}{\partial \ln{T}} n_l \frac{d T}{d x},\label{eq: thermal-est-1}
\end{align}
where $\lambda = v_{th}/\nu_{lh}$ is used to obtain the right side. 

Next, to contrast thermo-diffusion in neutral mixtures and plasmas with the
help of Eq.~(\ref{eq: thermal-est-1}), we consider an ensemble of
particles interacting via the potential
$\phi(r) \propto 1/r^n$. 
By balancing the kinetic and potential energies, the closest
approach $d$ between particles is found to scale as 
$ d \propto v^{-2/n}$
, where $v\propto
\sqrt{T}$ is the characteristic particles' velocities. For the
collision cross-section this gives $\sigma \propto d^2 \propto
T^{-2/n}$ and the collision frequency scaling can be recovered from
$\nu_{lh} \propto \left<\sigma v\right> $, where $\left<~\right>$ denotes the ensemble
average. As a result, one finds $\partial \ln{\nu_{lh}}/\partial\ln{T} = (n-4)/2n$.

%\begin{align}
%\frac{\partial \ln{\nu_{lh}}}{\partial\ln{T}} = \frac{n-4}{2n}.
%\end{align}

In a fully ionized plasma, $n=1$ so the factor on the right side of
Eq.~(\ref{eq: thermal-est-1}) is predicted to be $-3/2$. Interaction
between particles in neutral mixtures is limited to a shorter range.
For the purpose of the estimate in simple gases the corresponding $n$ can be
placed between $6$ and $12$~\cite{hirschfelder}, predicting the
factor on the right side of Eq.~(\ref{eq: thermal-est-1}) to lie
between $1/6$ and $1/3$. Hence, while in neutral mixtures
thermo-diffusion should indeed be substantially less significant than
baro-diffusion, the two should be comparable in plasmas with multiple
ion species. Equally important, the qualitative analysis also shows
that the thermal force by the heavy ion species on the light one,
$R_T^{lh}$, is in the direction opposite to the temperature gradient
($\partial \ln{\nu_{lh}} /\partial \ln{T} < 0$). Obviously, $R_T^{hl}
= -R_T^{lh}$ and therefore, as far as the relative motion of the two
species is concerned, the ion-ion thermal and $\nabla p_i$ forces act
together. It is thus predicted that, unlike the neutral mixture case
($\partial \ln{\nu_{lh}} /\partial \ln{T} > 0$), thermo-diffusion with
$\nabla T_i$ reinforces baro-diffusion in plasmas.

The thermal forces exerted onto the two ion species by electrons are
aligned and their combined effect on these species' relative motion is
not possible to predict with the above simplified analysis. As the rigorous
calculation will show, $ k_T^{(e)}$ and $k_p$ may have opposite signs,
but in such a case $ |k_T^{(e)}|$ is noticeably less than $|k_T^{(i)}|$ 
and the overall thermo-diffusion should reinforce
baro-diffusion still.
%Thermo-diffusion with $\nabla T_e$ can be in either direction....

{\bf Quantifying thermo-diffusion via first-principle calculation:}
We now proceed with first principle evaluation of the diffusion
coefficients in Eq.~(\ref{eq: canonical-flux}). The total collisional
force exerted on species $\alpha$ by other species is known to have
the form~\cite{hirshman-sigmar, zhdanov}
\begin{equation}
\label{eq: friction}
{\bf R}_{\alpha} =-\sum_{\beta} [A_{\alpha\beta}\mu_{\alpha\beta} n_{\alpha} \nu_{\alpha\beta} ({\bf u}_{\alpha} - {\bf u}_{\beta})+B_{\alpha\beta} n_{\beta}\nabla T_{\beta}] ,
\end{equation}
where $T_{\alpha}$ and $m_{\alpha}$ are, respectively, the temperature
and elementary mass of species $\alpha$; $\mu_{\alpha\beta}\equiv
m_{\alpha}m_{\beta}/(m_{\alpha}+m_{\beta})$ is the reduced mass and
$\nu_{\alpha\beta}$ stands for the frequency of collisions between
species ${\alpha}$ and ${\beta}$. Subscripts ``$\alpha$" and
``$\beta$" can take values ``$l$", ``$h$" and ``$e$" to denote light
ion, heavy ion and electron species, respectively. The first term on
the right side represents dynamic friction, and $ A_{\alpha\beta} =
A_{\beta\alpha}$ due to momentum conservation.  For realistic currents
the effect of ion-electron friction on mass diffusion can be neglected
($A_{he},A_{le} \approx 0$)~\cite{electro-diffusion}. The second term
on the right side represents thermal forces. The ion-electron thermal
force is governed solely by the electron temperature gradient
($B_{el},B_{eh} \approx 0$), because the thermal speed of electrons is
much greater than that of ions; both $B_{ll}$ and $B_{lh}$ ($B_{hh}$
and $B_{hl}$) contribute to the ion-ion thermal force.

Utilizing Eq.~(\ref{eq: friction}) to write the momentum conservation
equations for the two ion species and imposing the
short-mean-free-path ordering one can obtain the diffusive flux in the
form~(\ref{eq: canonical-flux}) from definition~(\ref{eq: flux-def})
with~\cite{electro-diffusion}
\begin{align}
\label{eq: diffusion-coeff}
&D= \frac{\rho T_i}{A_{lh}\mu_{lh} n_l \nu_{lh}} \times \frac{c(1-c)}{cm_h+(1-c)m_l}, \\
\label{eq: baro-diff-ratio}
&k_p = c(1-c)(m_h-m_l)\Bigl( \frac{c}{m_l} +  \frac{1-c}{m_h}    \Bigr),\\
\label{eq: electro-diff-ratio}
&k_E = m_lm_h c(1-c)  \Bigl( \frac{c}{m_l} +  \frac{1-c}{m_h}    \Bigr)  \Bigl( \frac{Z_l}{m_l} - \frac{Z_h}{m_h}  \Bigr),\\
\label{eq: thermo-diff-ratio-ion}
&k_T^{(i)} =m_lm_h  \Bigl( \frac{c}{m_l} +  \frac{1-c}{m_h}    \Bigr) \Bigl[ \frac{cB_{ll}}{m_l} +\frac{(1-c)B_{lh}}{m_h}  \Bigr],\\
&k_T^{(e)} = m_lm_h  \Bigl( \frac{c}{m_l} +  \frac{1-c}{m_h}    \Bigr) \Bigl[ \frac{cZ_l}{m_l} +\frac{(1-c)Z_h}{m_h}  \Bigr] 
\times \notag \\
\label{eq: thermo-diff-ratio-el}
&~~~~~~~~~[(1-c)B_{le} - cB_{he} ] \frac{T_e}{T_i},
\end{align}
where $Z_{\alpha}$ denotes the charge number of species $\alpha$. Importantly, for comparable $m_l$ and $m_h$ the time of equilibration \emph{within} an ion species is about the same as that of equilibration \emph{between} the species, making $T_l \approx T_h \equiv T_i$. Equilibration between electrons and ions requires a much longer time, so distinction between $T_i$ and $T_e$ should generally be kept. This said, the ion-ion collision frequency can be written as
\begin{equation}
\label{eq: coll-freq}
\nu_{lh} = \frac{4(2 \pi)^{1/2}}{3} \frac{n_{h} }{\mu_{lh}^{1/2} T^{3/2}}
 \Bigl(  \frac{Z_{l} Z_{h}e^2}{4\pi\varepsilon_0}  \Bigr)^{2}
 \ln{\Lambda},
\end{equation}
where $\ln{\Lambda}$ is the Coulomb logarithm.

%$\Lambda_{lh} = (12\pi \varepsilon_0/Z_{l} Z_{h}e^2) r_D$ and $r_D$ is the Debye radius.

It can be observed that the baro- and electro-diffusion ratios $k_p$
and $k_E$ are obtained without invoking a kinetic calculation,
i.e. they are thermodynamic quantities~\cite{landau-hydro,
  electro-diffusion}.  The classical diffusion coefficient $D$
involves the transport coefficient $A_{lh}$ defining dynamic friction
between the ion species according to Eq.~(\ref{eq: friction}). This
has a clear physical interpretation: it is the friction that prevents the
species from running apart, so larger $A_{lh}$ should result in
smaller $D$, as it is indeed predicted by Eq.~(\ref{eq:
  diffusion-coeff}). Thermo-diffusion ratios $k_T^{(i)}$ and
$k_T^{(e)}$ involve $B_{ll}$ and $B_{lh}$ and $B_{le}$ and $B_{he}$,
respectively. According to Eq.~(\ref{eq: friction}), these
coefficients govern the ion-ion and ion-electron thermal forces,
respectively; Eqs.~(\ref{eq: thermo-diff-ratio-ion})-(\ref{eq:
  thermo-diff-ratio-el}) therefore agree with the intuitive picture
presented earlier in this Letter. The fact that $A_{\alpha\beta}$ and
$B_{\alpha\beta}$ enter Eqs.~(\ref{eq: diffusion-coeff}), (\ref{eq:
  thermo-diff-ratio-ion})~and~(\ref{eq: thermo-diff-ratio-el}) suggests
that $D$, $k_T^{(i)}$ and $k_T^{(e)}$ are non-thermodynamic; deviation
of the species' distribution functions from Maxwellian needs to be
accounted for when evaluating these coefficients.

Hence, we are led to conduct a kinetic calculation to obtain
$A_{\alpha\beta}$ and $B_{\alpha\beta}$.  To this end, the
monograph by Zhdanov~\cite{zhdanov} turns out to be
instrumental. Ref.~\cite{zhdanov} employs the Grad 21N moment method, where
kinetic equation for each species is replaced with a set of
moment equations of this species' distribution function,
which is truncated to keep the first 21 moments. Importantly, when
evaluating transport in neutral gases, keeping 13 moments per species
is sufficient, whereas to achieve satisfactory precision of transport
calculation in unmagnetized plasmas (or longitudinal transport
calculation in magnetized plasmas) keeping 21 moments is
necessary. This is a mathematical manifestation of the peculiarity of
Coulomb collisions, as compared to the shorter range collisions in
neutral gases, that was explained in physical terms earlier in this Letter.

\begin{figure}[htbp]
\includegraphics[width=0.30\textwidth,keepaspectratio]{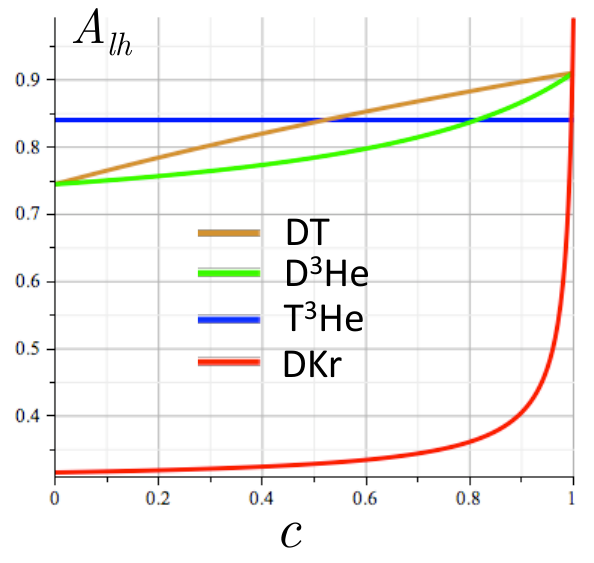}
\caption{The dynamic friction coefficient $A_{lh}$ as a function of $c$.}
\label{fig: A-lh}
\end{figure}

\begin{figure}[htbp]
\includegraphics[width=0.40\textwidth,keepaspectratio]{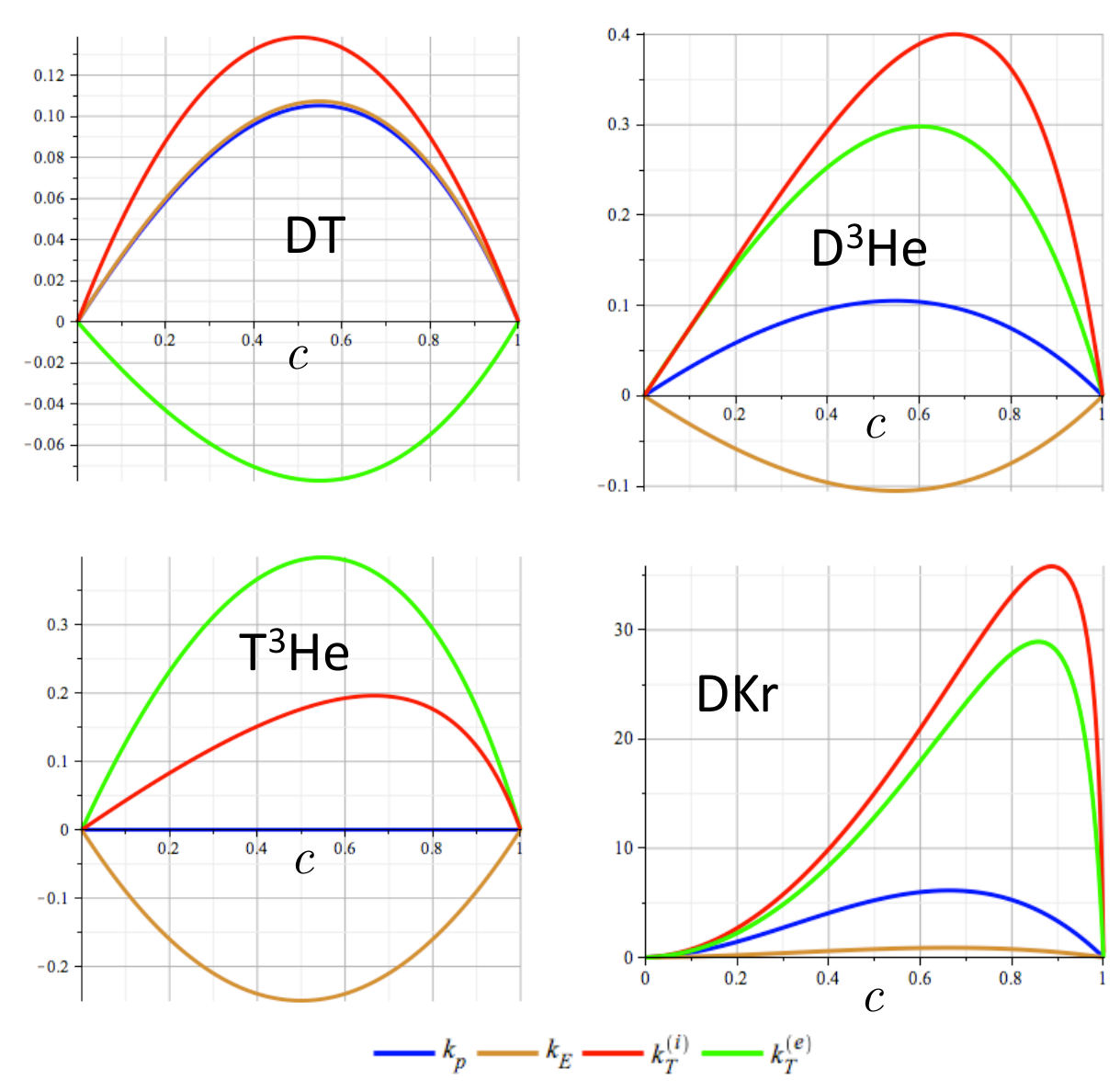}
\caption{Diffusion ratios $k_p$, $k_E$, $k_T^{(i)}$ and $k_T^{(e)}$ as functions of $c$.}
\label{fig: ks}
\end{figure}

By solving the short-mean-free-path version of the moment equations of
Ref.~\cite{zhdanov}, $A_{\alpha\beta}$ and $B_{\alpha\beta}$ as
functions of the species concentrations, elementary masses and charge
numbers can be obtained.  When $m_l$ and $m_h$ are comparable the
analytical  expression for $A_{lh}$, $B_{ll}$ and $B_{lh}$ becomes cumbersome,
so we numerically evaluate dependence of these coefficients on the
light species mass concentration $c$ for selected pairs of ion
species. The dynamic friction coefficient $A_{lh}$ as a function of
$c$ is plotted in Fig.~\ref{fig: A-lh}. Using numerically obtained $B_{ll}(c)$ and
$B_{lh}(c)$ along with Eq.~(\ref{eq: thermo-diff-ratio-ion}),
$k_T^{(i)}(c)$ is evaluated and plotted in Fig.~\ref{fig: ks}.

Since $m_e \ll m_{l,h}$ moment equations for electrons can be
separated from those for the two ion species~\cite{zhdanov}. This
allows tractable expressions for $B_{le}$ and $B_{he}$,
which, upon inserting into Eq.~(\ref{eq: thermo-diff-ratio-el}), give
\begin{equation}
\label{eq: thermo-diff-ratio-el-1}
k_T^{(e)} = - m_lm_h c(1-c) \Bigl( \frac{c}{m_l} +  \frac{1-c}{m_h}    \Bigr) \Bigl( \frac{Z_l^2}{m_l}  - \frac{Z_h^2}{m_h}  \Bigr)  
\frac{T_e  }{T_i } \frac{B_{ee}}{Z_{\bf eff}},
\end{equation}
where $Z_{\bf eff} \equiv (n_{l}Z_{l}^2 + n_{h}Z_{h}^2)/n_e $ is the
effective ion charge number, $n_{\alpha}$ the number density of
species $\alpha$, and
\begin{equation}
\label{eq: el-thermal-1}
B_{ee} \approx \frac{0.47+0.94Z_{\bf eff}^{-1}}{0.31+1.20Z_{\bf eff}^{-1}+0.41Z_{\bf eff}^{-2}}.                    
\end{equation}
Eqs.~(\ref{eq: thermo-diff-ratio-el-1})-(\ref{eq: el-thermal-1}) give
the $k_T^{(e)}(c)$ dependence for any given $m_{l,h}$ and $Z_{l,h}$;
this dependence is plotted in Fig.~\ref{fig: ks} for the pairs of
ion species, for which $k_T^{(i)}(c)$ is evaluated numerically.

%When solving for $A_{\alpha\beta}$ and $B_{\alpha\beta}$ as described,
%the species' heat conductivities are also recovered. These results are
%outside the scope of this Letter and will be presented elsewhere.

Eq.~(\ref{eq: canonical-flux}), along with Eqs.~(\ref{eq:
  baro-diff-ratio}),~(\ref{eq: electro-diff-ratio}) and (\ref{eq:
  thermo-diff-ratio-el-1}) and data in Figs.~\ref{fig: A-lh} and~\ref{fig: ks}, provide a
first-principle based description of mass diffusion in a plasma with
two ion species. With this result in hand, we are in a position to
evaluate the practical implications of this phenomenon in ICF plasmas.

It should first be observed in Fig. 1 that when ion masses are close,
kinetic effects do not introduce a substantial $c$ dependence in
$D$. In particular, for the T\textsuperscript{3}He mix, where $m_l =
m_h$, $A_{lh}(c)$ is perfectly flat. However, as the mass ratio
becomes larger, $A_{lh}$ may vary with concentration by a noticeable
factor, which is about three for the DKr mix ($m_h/m_l \approx 42$).

Second, in agreement with the qualitative analysis presented earlier in
this Letter, thermo-diffusion due to ion temperature gradient always
reinforces baro-diffusion. In contrast with  the
neutral mixture theory, 
baro-diffusion by no means dominates over
thermo-diffusion. Instead, for low-$Z$ mixtures with comparable masses
$k_T^{(i)}$ is slightly larger than $k_p$. For the DT mix, $k_T^{(e)}$
is negative, but small and is unlikely to substantially affect the
mass flux. 

Finally, in capsules doped with high-$Z$ impurities such as
Kr, it is thermo-diffusion, with both $\nabla T_i$ and $\nabla T_e$,
that dominates over baro-diffusion!

\begin{comment}
  Finally, the absolute values of $k_T^{(i)}$ and $k_T^{(e)}$ becoming
  large in the doped case (up to $\sim30$ for DKr) may explain the
  extreme sensitivity of the yield to the dopant amount observed
  experimentally~\cite{wilson-premix, garbett-premix,
    dodd-laser-absorp}. Indeed, according to equation~(\ref{eq:
    canonical-flux}) the concentration gradient should grow in
  response to externally imposed gradients of the pressure,
  temperature or electric potential. For a given background scale,
  i.e. given $\nabla \log{(p,T)}$, the larger diffusion ratios should
  therefore result in larger concentration perturbation.
\end{comment}

{\bf Mitigating fuel stratification:} In principle, mitigating the
negative consequence of inter-ion-species diffusion for ICF does
not necessarily require suppressing the diffusive flux at every point
in time in space throughout the implosion. One could attempt to
balance the concentration perturbations at different stages of the
implosion so that by the burn time the concentration breakdown becomes
the optimal one. However, this path requires a very careful timing of
the implosion and the diffusion effect may be difficult to untangle
from other potential sources of discrepancy between single-fluid
simulations and the experiment. For a more robust verification of the effect, it is desirable to find a case with ${\bf i}
\equiv 0$ and $c = const.$

\begin{comment}
Moreover, in exploding pushers experiments, the yield mostly comes from the region of high background gradient and mitigating the instantaneous diffusive flux is likely to be the only way to avoid this yield's degradation due to  species separation.
\end{comment}

Considering the adiabatic case with $T_e = T_i \equiv T$ and setting
the right side of Eq.~(\ref{eq: canonical-flux}) equal zero gives a
solubility constraint in terms of the diffusion ratios
\begin{equation}
\label{eq: adiabat}
1-\gamma = \frac{k_p + k_E}{k_p + k_E + k_T^{(i)} + k_T^{(e)} + B_{ee} k_E}.
\end{equation}
 With the
results of Fig. 2 this constraint, in turn, gives an equation for $c$,
for which the diffusion vanishes. Using the numerically obtained
dependencies of the diffusion ratios on $c$ and taking $\gamma = 5/3$, we find that ${\bf i}
\equiv 0$ is not possible to achieve for the DT mix. In the
D\textsuperscript{3}He mix, electro-diffusion acts in the favorable
direction, but is still unable to cancel the combined effect of the
baro- and thermo- diffusions. However, the D\textsuperscript{3}He case
suggests that the desirable regime may be realizable for the pair of
species with larger charge-to-mass difference, and therefore stronger electro-diffusion, such as
T\textsuperscript{3}He. This is confirmed by a numerical calculation,
which shows that for the T\textsuperscript{3}He diffusion ratios Eq.~(\ref{eq: adiabat}) does have a solution at $c\approx 0.89$.

In conclusion, first-principle derivation of the diffusive mass flux
in a plasma with two ion species reveals the paramount importance of
thermo-diffusion.  This is fundamentally due to the long range nature
of Coulomb potentials, in sharp contrast to short range collisions in
neutral gases.  Specifically, thermo-diffusion is found to contribute
to species separation as much as baro-diffusion in low-Z ion mixtures
and dominate over baro-diffusion in plasmas with high-Z dopants.
Since the thermo-diffusion ratios are much greater than unity in the
presence of a high-Z ion species, thermo-diffusion has the potential
to drive significant species separation when temperature gradient is large and
may contribute to
the aggravated yield anomaly in the doped capsule experiments.
We also show that diffusionless implosion is possible in
T\textsuperscript{3}He capsule with the concentration of T about
$c_* \approx 0.89$.  Assuming that other mechanisms potentially responsible for the yield anomaly are less sensitive to $c$, the role of the ion concentration diffusion on the performance
of ICF capsules can thus be studied by performing implosions of
T\textsuperscript{3}He filled targets with various concentrations of T.
The  discrepancy between the experimentally observed yield of the reaction
products with that predicted by standard single-fluid rad-hydro codes  should then
experience a minimum for $c \approx c_*$.

\begin{acknowledgements}

  The authors are indebted to the members of the MIT HEDP group for their many
  insights into experimental implications of the inter-ion-species
  diffusion. They also wish to thank F. J. Wysocki, G.A. Kyrala,
  Y.H. Kim and H.W. Herrmann of LANL and P. Helander of IPP-Greifswald for useful discussions.  This
  work was supported by the Laboratory Directed Research and
  Development (LDRD) program of LANL.

\end{acknowledgements}


\begin{thebibliography}{10}

%\bibitem{note-yield} If instead the laser driver rather than the target temperature is held fixed,
%optimal fusion yield can be obtained with different reactant number densities
%which leads to a higher target temperature upon implosion, especially in sub-ignition plasmas.

\bibitem{ICF} J. Nuckolls, L. Wood, A. Thiessen, and G. Zimmerman, Nature (London) 239, 139 (1972).

\bibitem{C-K-Li-mix} C. K. Li, F. H. SŽguin, J. A. Frenje, S. Kurebayashi, R. D. Petrasso, D. D. Meyerhofer, J. M. Soures et al, Phys. Rev. Lett. {\bf 89}, 165002 (2002).

\bibitem{wilson-mix-2011} D. C. Wilson, P. S. Ebey, T. C. Sangster, W. T. Shmayda, and R. A. Lerche, Phys. Plasmas {\bf 18}, 112707 (2011).


\bibitem{regan-Ar} S. P. Regan et al., Phys. Plasmas, {\bf 9}, 1357 (2002).

\bibitem{yaakobi-Kr} B. Yaakobi, R. Epstein, C. F. Hooper, Jr., D. A. Haynes, Jr, and Q. Su,  J. of X-ray Sci. and Tech. {\bf 6}, 172 (1996).

\bibitem{wilson-premix} D. C. Wilson, G. A. Kyrala, J. F. Benage Jr, F. J. Wysocki, M. A. Gunderson, W. J. Garbett, V. Yu Glebov et al, J. Phys.: Conf. Ser.  {\bf 112},  022015 
(2008).

\bibitem{garbett-premix} W. J. Garbett, S. James, G. A. Kyrala, D. C. Wilson, J. F. Benage, F. J. Wysocki, M. A. Gunderson, J. Frenje, R. Petrasso, V. Yu. Glebov, and B. Yaakobi, J. Phys.: Conf. Ser. {\bf 112}, 022016 (2008).

\bibitem{amendt-baro} P.~Amendt, O.~L.~Landen, H.~F.~Robey, C.~K.~Li and R.~D.~Petrasso, Phys. Rev. Lett. {\bf 105}, 115005 (2010).

\bibitem{zeldovich} Ya.~B.~Zel'dovich and Yu.~P.~Raizer,  Physics of Shock Waves and High-Temperature Hydrodynamic Phenomena (Dover, Mineola, New York, 2002).


\bibitem{landau-hydro} L.~D.~Landau and E.M.~ Lifshitz,  Fluid Mechanics (Oxford, Pergamon, 1958).

\bibitem{casey-prl} D.~T.~Casey, J.~A.~Frenje,  M.~G.~Johnson,  M.~J.~E.~Manuel,  H.~G.~Rinderknecht,  N.~Sinenian,  F.~H.~S{\'e}guin,  and C.~K.~Li,  R.~D.~Petrasso et al,
Phys. Rev. Lett. {\bf 108}, 075002 (2012).



\bibitem{Larroche-2012} O. Larroche, Phys. Plasmas {\bf 19}, 122706 (2012).

\bibitem{bellei-2013} C. Bellei, P. A. Amendt, S. C. Wilks, M. G. Haines, D. T. Casey, C.K. Li, R. Petrasso, and D. R. Welch, Phys. Plasmas {\bf 20}, 012701 (2013).

\bibitem{Larroche-comment} O. Larroche, Phys. Plasmas {\bf 20}, 044701 (2013).

\bibitem{bellei-reply-to-Larroche} C. Bellei, P. A. Amendt, S. C. Wilks, M. G. Haines, D. T. Casey, C.K. Li, R. Petrasso, and D. R. Welch, Phys. Plasmas {\bf 20}, 044702 (2013).




\bibitem{dodd-laser-absorp} E. S. Dodd, J. F. Benage, G. A. Kyrala, D. C. Wilson, F. J. Wysocki, W. Seka, V. Yu Glebov, C. Stoeckl, and J. A. Frenje, Phys. Plasmas {\bf 19}, 042703 (2012).

\bibitem{electro-diffusion} G.~Kagan and X.Z.~Tang,  Phys. Plasmas {\bf 19}, 082709 (2012).

\bibitem{amendt-lapse-rate} P.~Amendt, C.~Bellei and S.C.~Wilks,  Phys. Rev. Lett. {\bf 109}, 075002 (2012).

\bibitem{amendt-lapse-rate-reply} P.~Amendt, C.~Bellei and S.C.~Wilks,  Phys. Rev. Lett. {\bf 109}, 269502 (2012).


%\bibitem{amendt-pop} P.~Amendt, S.~C.~Wilks, C.~Bellei, C.~K.~Li and R.~D.~Petrasso, Phys. Plasmas {\bf 18}, 056308 (2011).

%\bibitem{herrmann-2009} H.~W.~Herrmann et al, Phys. Plasmas {\bf 16}, 056312 (2009).


%\bibitem{garbett-2008} W.~J.~Garbett et al, J. Phys.: Conf. Ser. {\bf 112}, 022016 (2008).

%\bibitem{lindl-2004} J.~D.~Lindl et al, Phys. Plasmas 11, 339 (2004)

\bibitem{braginskii} S.~I.~ Braginskii, Reviews of Plasma Physics {\bf 1}, 205 (1965).

\bibitem{hirschfelder} J.~O.~Hirschfelder, C.~F.~Curtiss and R.~B.~Bird, Molecular theory of gases and liquids (Vol. 26) (New York, Wiley, 1954).


\bibitem{zhdanov} V. M. Zhdanov, Transport Processes in Multicomponent Plasma, Taylor and
Francis, New York, (2002).

\bibitem{hirshman-sigmar} S.~Hirshman and D.~Sigmar, Nucl. Fusion {\bf 21}, 1079 (1981).


%\bibitem{note-grad} For the  adiabatic equation to be valid the background profiles should not be too sharp. In exploding pusher implosions most yield comes from the rebound shock, which is relatively weak, and this assumption is satisfactorily satisfied.

%\bibitem{rygg-science} J.~R.~Rygg, F.~H.~S{\'e}guin, C.~K.~Li,  J.~A.~Frenje,  M~.J.~E.~Manuel,  R.~D.~Petrasso, R.~Betti,  
%J.~A.~Delettrez,  O.~V.~Gotchev and  J.~P.~Knauer,  Science {\bf 319}, 1223 (2008).





\end{thebibliography}
\end{document}